\documentclass[preprint,aps,12pt,showpacs,nofootinbib,tightenlines]{revtex4-1}
\usepackage{amsmath}
\usepackage{amssymb}
\usepackage{epsfig}
\usepackage{dcolumn}
\usepackage{graphicx}
\usepackage{color}
\newcommand{\beq}{\begin{eqnarray}}
\newcommand{\eeq}{\end{eqnarray}}
\newcommand{\non}{\nonumber\\ }

\def \cpc{ {\bf Chin. Phys. C} }

\def \epjc{{\bf Eur.Phys.J. C} }
\def \epl{{\bf Europhys. Lett.} }
\def \scp{{\bf Sci China Phys Mech $\&$ Astron} }
\def \csb{ {\bf Chin. Sci. Bull.} }

\def \npb{ {\bf Nucl.Phys. B} }
\def \plb{ {\bf Phys.Lett. B} }
\def \pr{  {\bf Phys. Rept.} }
\def \prd{ {\bf Phys. Rev. D} }
\def \prl{ {\bf Phys.Rev.Lett.}  }

\def \jhep{ {\bf JHEP}  }
\def \jcap{ {\bf JCAP} }
\def \cpct{ {\bf Comput.Phys.Commun.} }

\begin{document}
\title{Constraining dark matter in the LRTH model with latest LHC, XENON100 and LUX data}
\author{Yao-Bei Liu$^{1,2}$, Zhen-Jun Xiao$^{1,3}$\footnote{Email: xiaozhenjun@njnu.edu.cn} }
\affiliation{1. Department of Physics and Institute of Theoretical Physics,
Nanjing Normal University, Nanjing 210023, P.R.China }
\affiliation{2. Henan Institute of Science and Technology, Xinxiang 453003, P.R.China}
\affiliation{3. Jiangsu Key Laboratory for Numerical Simulation of Large Scale
Complex Systems, Nanjing Normal University, Nanjing 210023, P.R. China}

\date{\today}
\begin{abstract}
In the left-right twin Higgs (LRTH) model, the neutral $\hat{S}$ is a candidate for weakly interacting
massive particle (WIMP) dark matter (DM). If its mass is lighter than half of the SM-like Higgs
boson $h$, the new invisible decay $h \to \hat{S}\hat{S}$ will become open.
In this paper, we examine the status of a light dark matter ( $\hat{S}$ ) under current experimental
constraints including the latest LHC Higgs data, the XENON100 and LUX limit on the dark matter scattering
off the nucleon.  The following observations have been obtained:
(i) The current ATLAS (CMS) measurements of $R_{\gamma\gamma}$
can exclude the invisible Higgs branching ratio ${\rm Br}_{\rm inv}$ larger than  $34\%$ ($48\%$) at $2\sigma$ level;
(ii) the Global fits to the latest LHC and Tevatron Higgs data provide a stronger constraint:
${\rm Br}_{\rm inv}< 20\%$ ($30\%$) at $2\sigma$ ($3\sigma$) level, which could be tested in the LHC experiments;
(iii) for the spin-independent scattering cross section off the nucleon, the recent XENON100 (LUX) data can exclude
the invisible decay rate larger than $50\%$ (25\%);
and (iv) the results of direct DM searches with LUX can give strong constraint on the viable  parameter
space of $\{ g_{h\hat{S}\hat{S}},  m_{\hat{S}}\}$ in this LRTH model.
\end{abstract}


\maketitle

\newpage
\section{Introduction}

To solve the little hierarchy problem in the standard model (SM), the twin Higgs mechanism \cite{ly-1,ly-2} is
proposed and the SM Higgs emerges as a pseudo-Goldstone boson
once a global symmetry is spontaneously broken. The left-right twin Higgs (LRTH) model is an economical
realization for this mechanism which is implemented in left-right models. Furthermore, an additional
discrete symmetry is introduced to render no quadratic divergence in the Higgs mass squared. The LRTH
model predicts several physical Higgs bosons in which the lightest particle $\hat{S}$ in the odd $SU(2)_L$
neutral components is stable, thus can be a good candidate for weakly interacting massive particle (WIMP)
dark matter (DM) \cite{ly-2}. The collider phenomenology of the LRTH model has been studied intensively for
example in Refs.~\cite{Hock,dong,dong1,dong2,sf,wl-jhep}.

From the theoretical point of view, the WIMP is one popular candidate of DM \cite{wimp}. It is interesting
to understand the viable WIMP mass range and its related couplings under current DM experimental constraints.
On the experimental side, various underground direct detection experiments, such as
DAMA, CoGeNT and CRESST, have found some DM-like events in the
low region \cite{dama}. Recently, the CDMS-II Collaboration \cite{cdms} reported that three WIMP-candidate
events were observed by using the silicon detectors. This observation, however, seems to be contradict with
the XENON100 data \cite{xenon} or the latest LUX data \cite{lux}, which
provided the most stringent upper limits on the spin-independent WIMP-nucleon scattering
cross section for a WIMP with a mass above 10 GeV. These rapid progress of direct DM detection experiments
allow a test for NP in the LRTH model.
The implications of the new results from the DM experiments have been widely
explored in Refs.~\cite{dm1,dm2,dm3}.

On the other hand, the discovery of a Higgs boson with a mass of roughly 125 GeV has been confirmed
by both the ATLAS \cite{atlas-1} and CMS \cite{cms-1} collaborations with the full RUN-1 data,
which is also supported by the CDF and D0 collaborations at the Tevatron \cite{tevatron}.
During the ICHEP 2014, some new results about the Higgs boson were presented by both ATLAS and CMS
Collaborations. Especially the signal strength of the diphoton decay channel of ATLAS has changed
from $1.6\pm0.3$ to $1.17\pm0.27$ \cite{atlas-2} and that of CMS from $0.77\pm0.27$ to $1.12^{+0.37}_{-0.32}$
\cite{cms-2}. The updated experimental measurements for $\mu_{\gamma\gamma}$
from both ATLAS and CMS become now well consistent with the SM prediction in 1$\sigma$ range,
and can put strong constraints on various NP models \cite{zhu}.
The SM-like Higgs boson properties in the LRTH model has been studied in our recent work \cite{liu-1311},
and the signal rates normalized to the corresponding SM predictions were found to be suppressed.

If the mass of the DM is less than half of the Higgs mass, the Higgs could decay into the light
DM pairs with a large invisible branching ratio ${\rm Br}_{\rm inv}$, which is very sensitive to the
NP \cite{lhc-inv}. The current strongest limits come from the CMS, which uses the vector boson fusion
and associated ZH production modes
to achieve a constraint of ${\rm Br}_{\rm inv}<0.58$ at $95\%$ CL \cite{cms-br}. Very recently,
N. Zhou et al \cite{tth} have shown that ${\rm Br}_{\rm inv}<0.4$ at $95\%$ confidence level (CL)
from $t\bar{t}h$ production. Recent results from the LHC and the direct DM detection experiments constrain considerably
possible scenarios beyond the SM \cite{npinv1,npinv2,npinv3}. This motivates us to explore
the implications for the DM in the LRTH model from those new experimental measurements.

In this work, we study the invisible decay mode $h \to \hat{S}\hat{S}$ of the SM-like
Higgs boson under the latest experimental constraints from the updated LHC Higgs data, the XENON100
and LUX limit on the DM-nucleon scattering. In addition, we also display the allowed parameter space
$\{g_{h\hat{S}\hat{S}}, m_{\hat{S}}\}$ in comparison
with the direct detection results of XENON100 and LUX for both low and high DM mass regions. In Sec.~II,
we  recapitulate the dark matter sector in the LRTH model. In Sec.~III, we give the
numerical results, and present the constraints on the parameter space in this model from the latest
experimental results. Finally, we summarize our conclusion in last section.

\section{The dark matter in the LRTH model}\label{sec:intro}

Here we will only focus on the dark matter sector in the LRTH model. For more details such as the
particle spectrum and the Feynman rules, one can refer to Ref.~\cite{Hock}.
The LRTH model contains two Higgs fields: $H=(H_{L},H_{R})$ and $\hat{H}=(\hat{H}_{L},\hat{H}_{R})$.
Under the gauge symmetry $SU(2)_{L}\times SU(2)_{R}\times
U(1)_{B-L}$, these scalars transform as
\beq H_{L}~and~ \hat{H}_{L}: (2, 1, 1),~~~~~~~~H_{R}~
and~ \hat{H}_{R}: (1, 2, 1).
\eeq
The global $U(4)_{1,2}$ symmetry is spontaneously broken down to its subgroup
$U(3)_{1,2}$ with vacuum expectation values (VEV) as: $<H>=(0,0,0,f)$
and $<(\hat{H})>=(0,0,0,\hat{f})$, which also break $SU(2)_{R}\times
U(1)_{B-L}$ down to the SM $U(1)_{Y}$. Each spontaneously symmetry
breaking yields seven Nambu-Goldstone bosons. After spontaneous global symmetry breaking and electroweak symmetry breaking, six out of the 14 Goldstone bosons are respectively eaten by the heavy gauge bosons $W^{\pm}_{H}$ and $Z_{H}$, and the SM gauge bosons
$W^{\pm}$, $Z$. There are left with one SM-like physical
Higgs boson $h$, a neutral pseudoscalar $\phi^{0}$, a pair of charged scalar $\phi^{\pm}$,
and an odd $SU(2)_{L}$ doublet $\hat{h}=(\hat{h}_{1}^{+},\hat{h}_{2}^{0})$. Here the lightest particle
in $\hat{h}$ is stable and thus can be treated as a candidate for dark matter.

In the LRTH model, the soft left-right symmetry breaking terms, so
called ``$\mu$-term", can give masse for $\hat{h}_2^0$ \cite{Hock}:
\beq
V_{\mu}=-\mu^2_r(H^\dagger_R\hat H_R+h.c.)+\hat \mu^2 \hat
H^\dagger_L \hat H_L. \label{muterm}
\eeq
Here $\mu_r$ should be smaller than
the value of $f/4\pi$ in order not to reintroduce fine tuning. Since $\hat{\mu}^2$ could have either sign, we can take the
masse of $\hat{h}_2^0$ as a free parameter.

The complex scalar $\hat{h}_2^0$ can be written as $\hat{h}_2^0=(\hat{S}+i\hat{A})/\sqrt{2}$,
where $\hat{S}$ and $\hat{A}$ are the scalar and pseudoscalar
fields, respectively. The mass splitting between $\hat{S}$ and $\hat{A}$ can be obtained by introducing a new
quartic potential term \cite{sf}:
\beq
V_{H}=-\frac{\lambda_{5}}{2}[(H^\dagger_L\hat H_L)^{2}+h.c.]. \label{lam5}
\eeq
Once $H_{L}$ obtains a VEV $(0, v/\sqrt{2})$, we can get a mass splitting between $m^{2}_{\hat{S}}$ and $m^{2}_{\hat{A}}$
\beq
m^{2}_{\hat{A}}-m^{2}_{\hat{S}}=\lambda_5v^2.
\label{del21}
\eeq
Here $\hat{S}$ is lighter than $\hat{A}$, and can be a candidate of DM. Note that the one-loop quadratic divergence of Higgs mass
from the $\hat{S}$ loop and from the $\hat{A}$ loop can be
cancelled due to the opposite sign in Eq. (\ref{lam5}).
From above equation, we can get the Higgs coupling $h\hat{S}\hat{S}$ and $h\hat{A}\hat{A}$ which are related to the
parameter $\lambda_5$. Similarly, the mass splitting between $m^{2}_{\hat{h}_{1}}$ and $m^{2}_{\hat{h}_{2}}$ can be generated by introducing a quartic term $V_{H}=-\lambda_{4}|H^\dagger_L\hat H_L|^{2}$\cite{sf}:
\beq
m^{2}_{\hat{h}_{1}}-m^{2}_{\hat{h}_{2}}=-\frac{\lambda_4v^2}{2}. \label{lam4}
\eeq
However, the Coleman-Weinberg (CW) potential can also give the contributions to the
couplings of $h\hat{S}\hat{S}$, $h\hat{A}\hat{A}$ and $h\hat{h}_{1}^{+}\hat{h}_{1}^{-}$. The exact expressions of these couplings are listed in Appendix A. Here we take a new free parameter $g_{h\hat{S}\hat{S}}$ as the coupling strength of
$h\hat{S}\hat{S}$.

The relic density analysis of the DM in the LRTH
 model has been studied in Ref.~\cite{sf}, which shows that the low DM mass
region  can satisfy the constraints of $\Gamma_{Z}$ and WMAP $3 \sigma$ relic
density.  Here we focus on such low mass region where the invisible decay
$h\to \hat{S}\hat{S}$ are open,
whose partial width is given by
\beq
\Gamma(h \to \hat{S}\hat{S}) =
 \frac{g_{h\hat{S}\hat{S}}^2}{8\pi m_h}\sqrt{1-\frac{4m_{\hat{S}}^{2}}{m_{h}^{2}}},
\eeq
Of course, it can also be determined by the branching ratio of the invisible
Higgs decay ${\rm Br}_{\rm inv}$, the light DM mass $m_{\hat{S}}$ and other
model parameters including $f$ and $M$. Note that the new decay modes $h\to
\hat{A}\hat{A}$ and $h\to\hat{h}_1^{+}\hat{h}_1^{-}$ can also be open for low
values of $m_{\hat{A}}$ and $m_{\hat{h}_{1}}$, but their decay branching
ratios are relatively small and can be neglected \cite{wl-jhep}.

\section{Numerical results and discussions}

In our calculations, we take the SM-like Higgs mass as
$m_{h}=125.5$ GeV. The SM input parameters relevant in our study are
taken from \cite{data}. The free parameters involved are $f$, $M$
and the invisible decay branching ratio ${\rm Br}_{\rm inv}$ (or the coupling
strength $g_{h\hat{S}\hat{S}}$ and the DM mass $m_{\hat{S}}$). The indirect constraints on $f$ come
from the $Z$-pole precision measurements, the low energy neutral current process and high energy precision
measurements off the $Z$-pole. The mixing parameter $M$ can be constrained by the
$Z\rightarrow b\bar{b}$ branching ratio and oblique parameters \cite{0701071}.
In Ref.~\cite{top-partner}, Atlas collaboration claimed that a $T$-partner with a masse below $656$ GeV are excluded
at $95\%$ confidence level  under the assumption of a branching ratio $BR(T\rightarrow W^{+}b)=1$.
In the LRTH model, however, the dominant decay mode for the heavy top partner is $T\rightarrow \phi^{+}b\rightarrow t\bar{b}b$
instead of the $T \to W^+ b$. The Atlas limit on $m_T$ as given in Ref.~\cite{top-partner} therefore
does not apply to our case.

Following Ref.~\cite{Hock}, we will assume that the values of the free parameters $f$ and $M$ are in the ranges of
\beq 500 \rm GeV \leq f \leq 1500 \rm GeV, \quad 0 \leq M
\leq 150 \rm GeV.
\eeq

\subsection{Implication of LHC Higgs data on the invisible decay}

For $m_{h}=125.5$ GeV, the production cross sections for each
production channels at the LHC could be found in Ref. \cite{handbook}. In the SM,
the dominant production process is $gg\rightarrow h$ by the top quark loop, while
the LRTH model can give corrections via the modified coupling of $h\bar{t}t$ and
the heavy T-quark loop. The hadronic cross section $\sigma (gg\rightarrow h)$ has
a strong correlation with the decay width $\Gamma (h\rightarrow gg)$. Thus, the
Higgs production diphoton rates
in the LRTH model normalized to the SM values can be defined as:
 \beq R_{\gamma\gamma}=\frac{[\Gamma(h \rightarrow
gg)\times {\rm Br}(h\rightarrow \gamma\gamma)]_{\rm LRTH}}{[\Gamma(h \rightarrow
gg)\times {\rm Br}(h\rightarrow \gamma\gamma)]_{\rm SM}},
\eeq
with
\beq
{\rm Br}(h\rightarrow \gamma\gamma)_{\rm LRTH}=\frac{\Gamma(h\rightarrow
\gamma\gamma)_{\rm LRTH}}{\Gamma_{\rm LRTH}+\Gamma(h \to \hat{S}\hat{S})}
=\frac{\Gamma(h\rightarrow \gamma\gamma)_{\rm LRTH}}{\Gamma_{\rm LRTH}}(1-{\rm Br}_{\rm inv}),
\eeq
where $\Gamma_{\rm LRTH}$ denotes the decay width of SM-like Higgs for
$m_{\hat{S}}> m_{h}/2$.  As mentioned in Appendix B, the decay width $\Gamma (h\rightarrow gg)_{\rm LRTH}$ and
$\Gamma (h\rightarrow \gamma\gamma)_{\rm LRTH}$ depend on the parameters $f$ and $M$.
The ratio $R_{\gamma\gamma}$ can therefore be determined by three parameters $f$, $M$,
and ${\rm Br}_{\rm inv}$.

In Fig.1 we plot $R_{\gamma\gamma}$ versus ${\rm Br}_{\rm inv}$ for $M=150$ GeV and $f=1000$ GeV, respectively. It can be
seen from Fig.1(a) that ratio $R_{\gamma\gamma}$ in the LRTH model is always smaller
than unit, and will approach one (the SM prediction) for ${\rm Br}_{\rm inv}=0$ and a large scale $f$. On
the other hand, we can see from Fig.1(b) that the ratio is insensitive to the
mixing parameter $M$. By using the improved measured values of $R_{\gamma\gamma}$ from both ATLAS and CMS
\beq
R_{\gamma\gamma}^{\rm exp}=\left\{ \begin{array}{ll}
1.17\pm 0.27 , & {\rm ATLAS}, \\
1.12^{+0.37}_{-0.32}, & {\rm CMS}, \\ \end{array} \right. \label{eq:r2gamma}
\eeq
we find that the predicted $R_{\gamma\gamma}$ based on the LRTH model is outside the
$1\sigma$ range of the ATLAS and CMS data in the most of the parameter space, as illustrated by Fig.~1(a).
For $f$=500 GeV, the predicted rate is  outside the $2\sigma$ range of the ATLAS data,
while the current CMS diphoton rate can exclude the invisible decay branching ratio about
$20\%$ at $2\sigma$ level.
In the reasonable parameter space, however, the theoretical prediction of
$R_{\gamma\gamma}$ can be in good agreement with the ATLAS (CMS) data in $2\sigma$ range if ${\rm Br}_{\rm inv} < 0.34~(0.48)$.

\begin{figure}[ht]
\begin{center}
\vspace{-0.3cm}
\centerline{\epsfxsize=9cm\epsffile{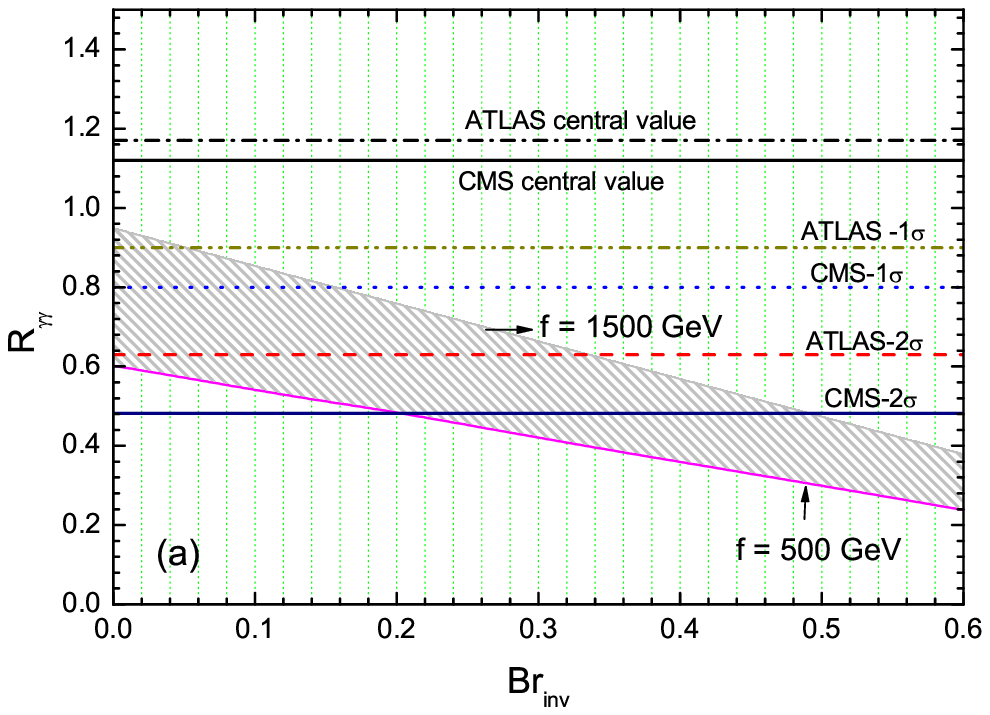}\hspace{-0.5cm}\epsfxsize=9cm\epsffile{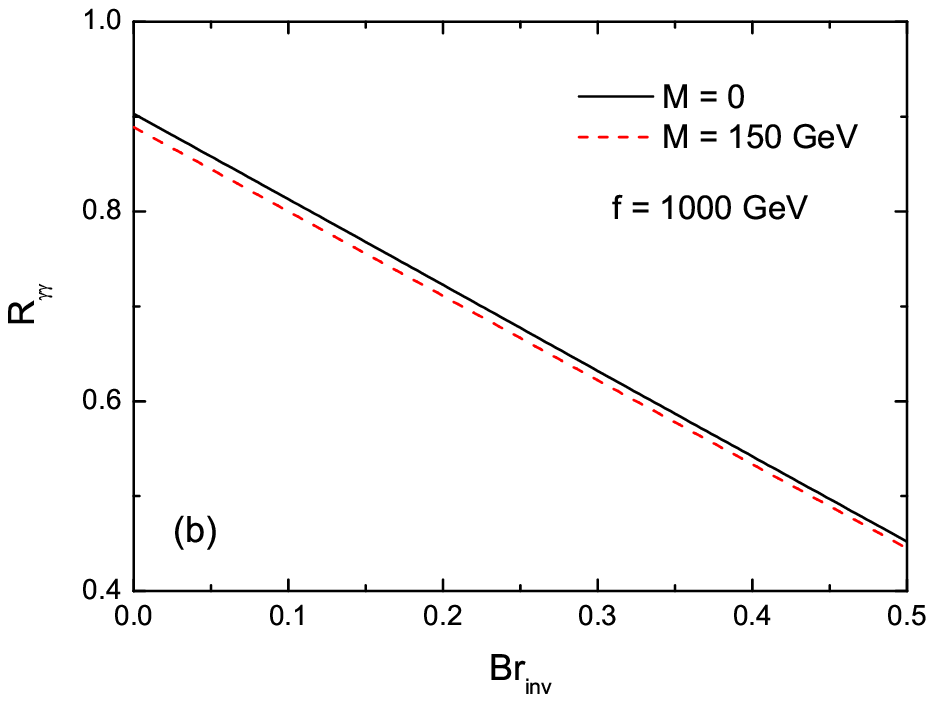}}
\caption{ ${\rm Br}_{inv}$-dependence of $R_{\gamma\gamma}$ for (a) $M=150$ GeV
and 500 GeV$\leq f\leq$1500 GeV, and (b) $f=1000$ GeV and $M=0$ and 150 GeV, respectively. }
\end{center}
\end{figure}

Now we perform a global fit to the LRTH model with the
method proposed in \cite{jhep1,jhep2,1407.8236} with the latest Higgs data as listed in Table I-V of
Ref.~\cite{1407.8236}. We will consider totally the 29 Higgs signal strength observables
from ATLAS, CMS, CDF and D0 collaborations for $\gamma\gamma$, $ZZ^{\ast}$, $WW^{\ast}$,
$b\bar{b}$, $\tau^{+}\tau^{-}$ and $t\bar{t}H$ channels as reported by the relevant collaborations
~\cite{atlas-2,atlas-3,atlas-4,cms-2,cms-3,tevatron}. In Table I, for the sake of the reader,
we show the latest Higgs data and the corresponding theoretical predictions in terms of
the LRTH parameters for the $\gamma\gamma$ channels only.

The global $\chi^{2}$ function is defined as:
\begin{equation}
\chi^{2}=\sum_{i,j}(\mu_{i}-\hat{\mu}_{i})(\sigma^{2})^{-1}_{ij}(\mu_{j}-\hat{\mu}_{j}),
 \end{equation}
where $\sigma^{2}_{ij}=\sigma_{i}\rho_{ij}\sigma_{j}$,
$\hat{\mu}_{i}$ and $\sigma$ are the measured Higgs signal strengths
and their $1\sigma$ error, $\rho_{ij}$ is the correlation matrix,
$\mu_{i}$ is the corresponding theoretical predictions in terms of the LRTH parameters.
The values of $\chi^{2}$ as listed in the last line of Table I are obtained after fitting for
all 29 Higgs  signal strength observables.

\begin{table}[htb]
\begin{center}
\caption{ The measured Higgs signal strengths and the LRTH predictions
for the $\gamma\gamma$ channels only (for other channels see Ref.\cite{1407.8236}).
Here we consider the cases for the fixed
${\rm Br}_{\rm inv}=0.1, 0.2, 0.3$, $M$=150 GeV and $f=1000$ GeV. } \label{table1}
\vspace{0.2cm}
\begin{tabular}{|c|c|ccc|} \hline \hline
Channel & Signal strength $\hat{\mu}_{i}$&\multicolumn{3}{c|}{LRTH predictions of $\mu_{i}$ }\\
 & &${\rm Br}_{\rm inv}=0.1\ \ $ & $\ \ 0.2\ \ $ &\hspace{0.4cm} $0.3 $\hspace{0.4cm}  \\ \hline
 \multicolumn{5}{|c|}{ATLAS \cite{atlas-2}}\\ \hline
ggF, $\gamma\gamma$&$1.32\pm0.38$ &0.845&0.751&0.657 \\
\hline
VBF, $\gamma\gamma$&$0.8\pm0.7$ &0.845&0.751&0.657 \\ \hline
WH, $\gamma\gamma$&$1.0\pm1.6$ &0.900&0.800&0.700 \\ \hline
ZH, $\gamma\gamma$&$0.1^{+3.7}_{-0.1}$ &0.900&0.800&0.700 \\ \hline
ttH, $\gamma\gamma$&$1.6^{+2.7}_{-1.8}$ &0.896&0.796&0.697 \\  \hline
\multicolumn{5}{|c|}{CMS \cite{cms-2}}\\ \hline
ggF, $\gamma\gamma$&$1.12^{+0.37}_{-0.32}$ &0.845&0.751&0.657 \\ \hline
VBF, $\gamma\gamma$&$1.58^{+0.77}_{-0.68}$ &0.900&0.800&0.700 \\ \hline
VH, $\gamma\gamma$&$-0.16^{+1.16}_{-0.79}$ &0.900&0.800&0.700 \\ \hline
ttH, $\gamma\gamma$&$2.69^{+2.51}_{-1.81}$ &0.896&0.796&0.697 \\  \hline
\hline \multicolumn{5}{|c|}{ $\cdots$ } \\ \hline \hline
$\chi^{2}$& 16.38&19.49&22.88&28.11 \\ \hline
\end{tabular} \end{center} \end{table}

\begin{figure}[ht]
\begin{center}
\vspace{-0.5cm}
\centerline{\epsfxsize=8cm\epsffile{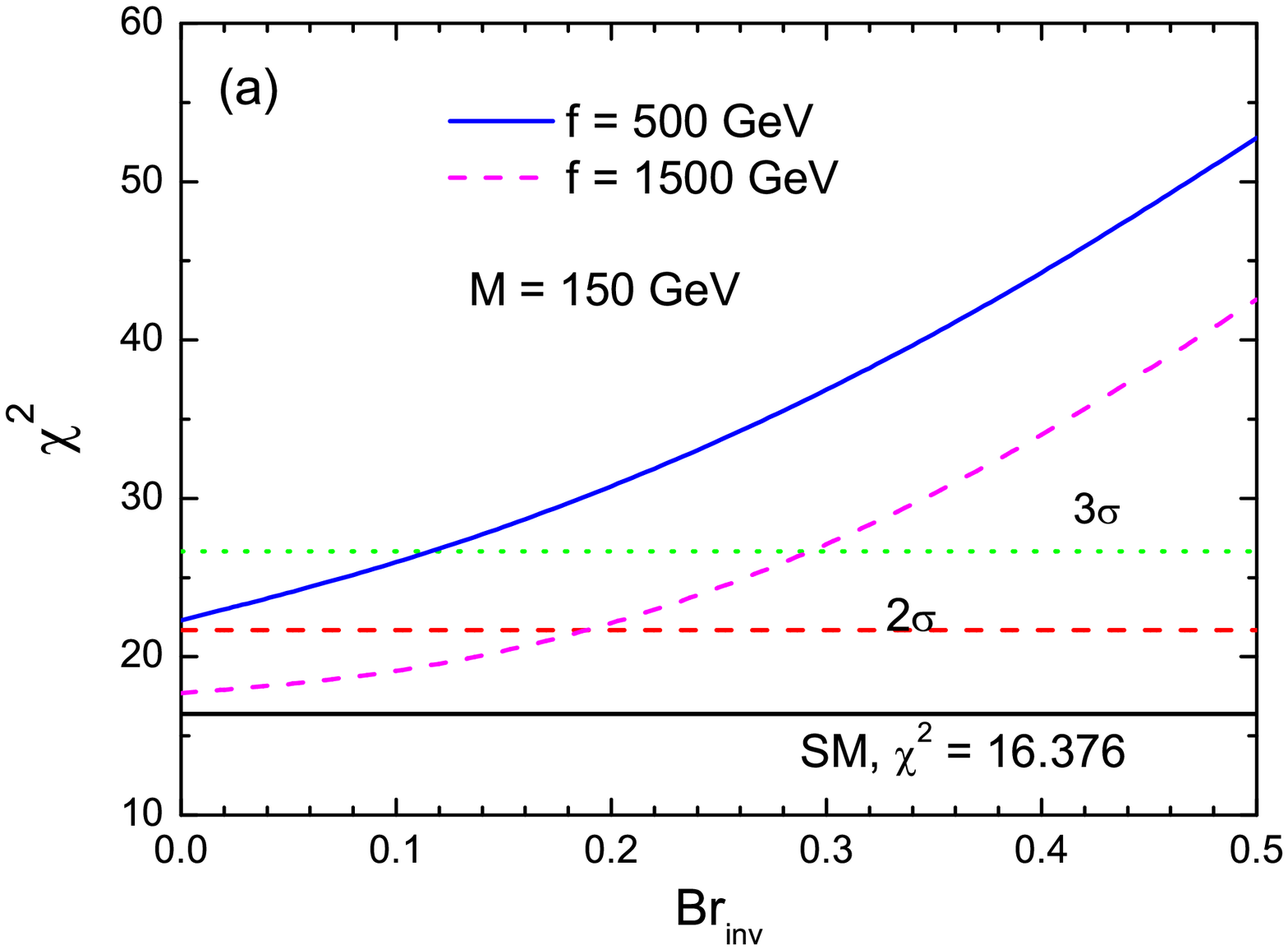}
\hspace{-0.5cm}\epsfxsize=8cm\epsffile{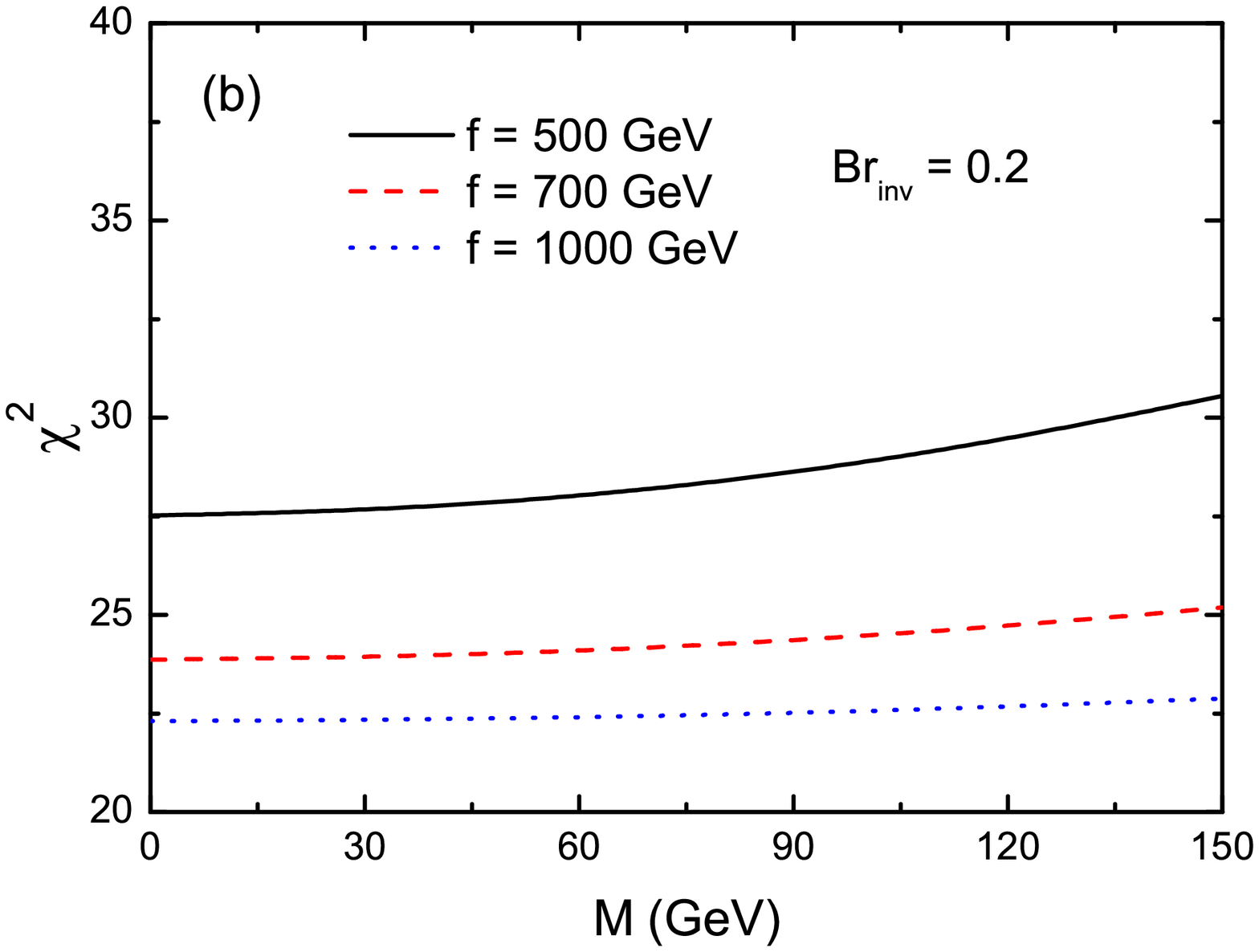}}
\caption{(a) The global fit values of $\chi^{2}$
versus ${\rm Br}_{\rm inv}$ for $M=150$ GeV and two values of $f$ as indicated;
(b) $\chi^{2}$ versus $M$ for ${\rm Br}_{\rm inv}=0.2$ and three values of $f$ as indicated.}
\end{center}
\end{figure}

In Fig.~2(a) we plot values of $\chi^{2}$ versus ${\rm Br}_{\rm inv}$ for $M=150$ GeV. One can see that the
value of $\chi^{2}$ is larger than that for SM for most of parameter
space and approaches the SM value for ${\rm Br}_{\rm inv}=0$. From Fig.2(b) one can
see that he value of $\chi^{2}$ is insensitive to the
mixing parameter $M$, and thus we can safely take $M=150$ GeV as the typical value in latter calculations.

In Fig.~3 we draw the contour plots of $\chi^{2}$ in $f-{\rm Br}_{\rm inv}$ plane.
One can also see that the current Higgs data can give strong constraint on the exotic decay
$h\to \hat{S}\hat{S}$: i.e., the ${\rm Br}_{\rm inv}$ should  be smaller
than $30\%$ ($20\%$) at 3$\sigma$ ($2\sigma$) level.
Such expectation could be tested at the LHC with $100 ~{\rm fb^{-1}}$ integrated
luminosity, where a $95\%$ C.L. upper limit on the invisible decay: ${\rm Br}_{\rm inv}< 0.18$
\cite{018}.
The $\chi^{2}$ at the minimum is $\chi^{2}_{\rm min}=17.67$, i.e. almost the same as for the SM.

\begin{figure}[t]
\begin{center}
\vspace{-0.5cm} \centerline{\epsfxsize=8cm\epsffile{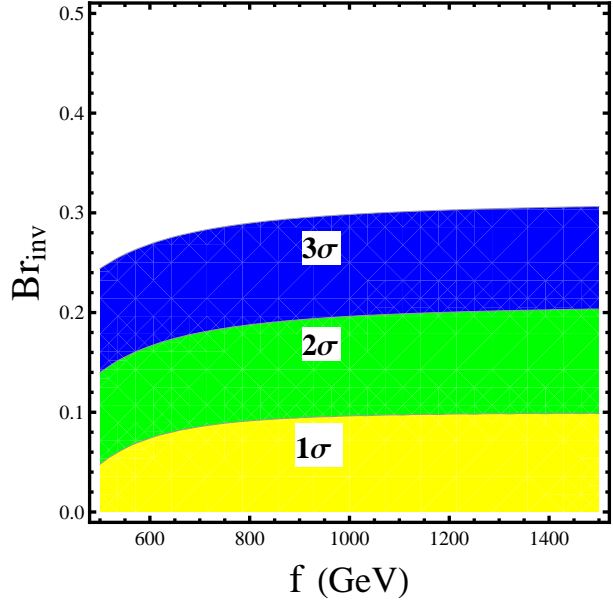}}
\caption{ The contours of $\chi^{2}$ in $f-{\rm Br}_{\rm inv}$ are shown for the
1$\sigma$, 2$\sigma$, and 3$\sigma$ regions. } \label{fig:fig3}
\end{center}
\end{figure}

\subsection{DM direct search experiments}

Very recently the stringent limits on the spin-independent component of elastic
scattering cross section $\sigma_{\rm SI}$ from XENON100
\cite{xenon} and LUX \cite{lux} experiments become available.
In LRTH model, the elastic scattering of $\hat{S}$ on a nucleus
receives the dominant contributions from the Higgs boson exchange
diagrams. Thus the LHC bounds on ${\rm Br}_{\rm inv}$ can be interpreted as the bounds
on the DM scattering off nucleons, mediated by Higgs exchange \cite{higgs-ex}.
The spin-independent elastic scattering cross section between $\hat{S}$ and
the nucleon is given by:
\begin{equation}
\sigma_{\rm SI}= \frac{2 m_r^2 m_p^2}{m_{h}^{3} m_{\hat{S}}^{2}\beta} \frac{g^2}{M_W^2}
\left (\frac{\Gamma_{\rm LRTH}\times{\rm Br}_{\rm inv}}{1-{\rm Br}_{\rm inv}} \right )
\left [ C_{U} (f_{u}^{N}+f_{c}^{N}+f_{t}^{N}) +C_{D} (f_{d}^{N}+f_{s}^{N}+f_{b}^{N})\right ]^2,
\label{sig1}
\end{equation}
where $\beta=\sqrt{1-(4m_{\hat{S}}^{2}/m_{h}^{2} ) }$, $m_{r}$ is the reduced mass and
$f_{q}^{N},(f_{g}^{N})$ are the quark (gluon) coefficients in the nucleon.
We take the values  $f_{s}^{p}=0.0447$, $f_{u}^{p}=0.0135$, and $f_{d}^{p}=0.0203$
from an average of recent lattice results~\cite{2013ac}. The gluon and heavy quark
($Q=c,b,t$) coefficients are related to those of light quarks,
and $f_{g}^{p}= f_{Q}^{p}= \frac{2}{27}(1-\sum_{q=u,d,s} f_{q}^{p})$ at leading order.

\begin{figure}[thb]
\begin{center}
\vspace{0.3cm}
\centerline{\epsfxsize=6.5cm\epsffile{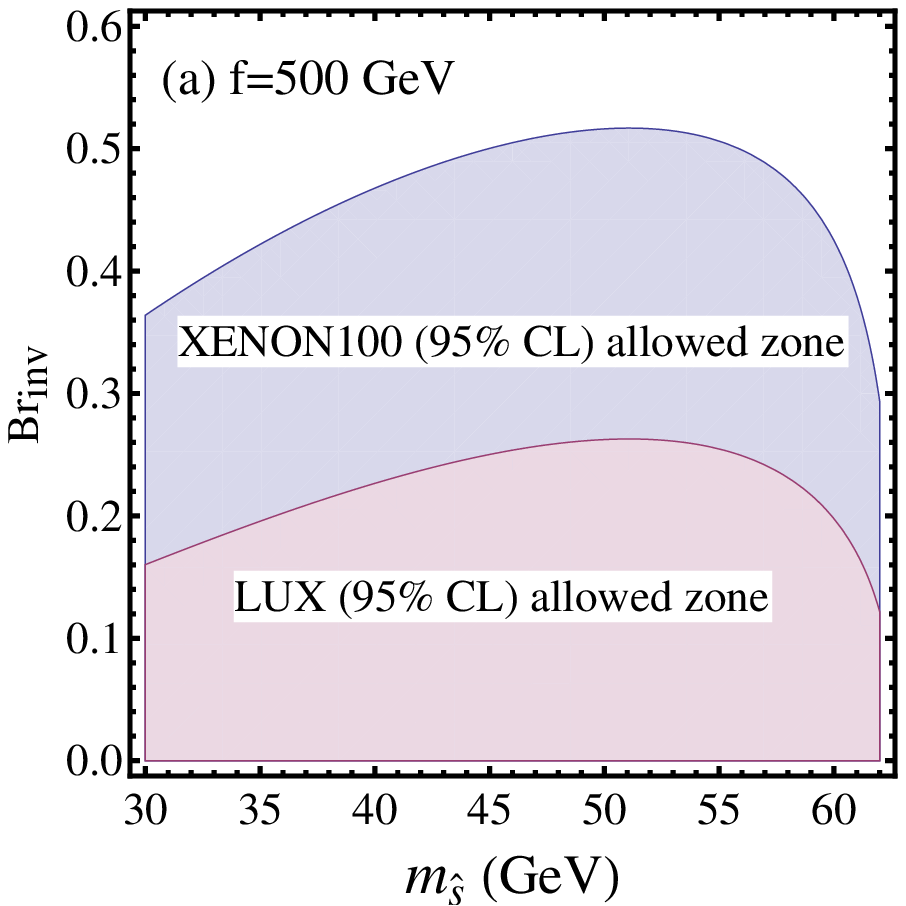}
\hspace{0.5cm}\epsfxsize=6.5cm\epsffile{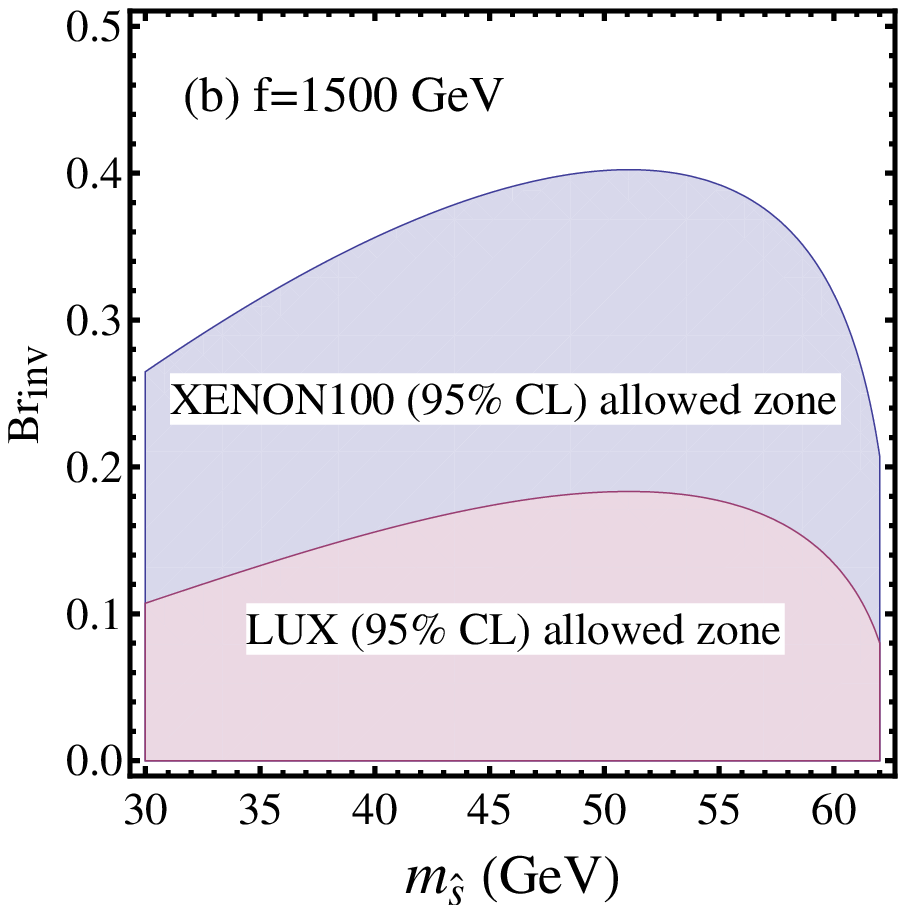}}
\caption{The parameter space allowed in the $m_{\hat{S}}-{\rm Br}_{\rm inv}$ plane
by the  XENON100 and LUX dates for $f=500$ GeV (left) and $f=1500$ GeV (right). }
\end{center} \end{figure}

For a given mass of the dark matter candidate, one can translate
the exclusion limit set by XENON100 and LUX experiments into an
upper bound on the invisible branching ratio of the SM-like Higgs boson.
The allowed parameter space of $m_{\hat{S}}-{\rm Br}_{\rm inv}$ by the experimental data
are displayed in Fig.~4.
One can see from Fig.~4 that the recent XENON100 data can exclude the region of $ {\rm Br}_{\rm inv} > 50\%$,
while the latest LUX result can exclude the region $ {\rm Br}_{\rm inv} > 25\%$.

\subsection{Bounds on the parameter space of $\{g_{h\hat{S}\hat{S}},  m_{\hat{S}}\}$}

From above discussions we get to know that the value of the coupling
$g_{h\hat{S}\hat{S}}$ could be determined for given values of ${\rm Br}_{\rm inv}$, the DM mass
$m_{\hat{S}}$ and the model parameters $f$ and $M$.
For a low DM mass region, we know that both the LHC and DM experimental results can give strong
bound on the two-dimensional parameter space  $\{g_{h\hat{S}\hat{S}}, m_{\hat{S}}\}$.
For a high DM mass region, on the other hand, the coupling $g_{h\hat{S}\hat{S}}$ can also be
strictly constrained by the XENON100 and LUX limits on the DM-nucleon scattering cross section.
Eq.~(\ref{sig1}) can also be written as the form of
\begin{equation}
\sigma_{\rm SI}= \frac{g^{2}m_r^2 \; m_p^2\; g^{2}_{h\hat{S}\hat{S}}}{4\pi M_{W}^{2}\; m^{4}_{h}m^{2}_{\hat{S}}}
\left [ C_{U} (f_{u}^{N}+f_{c}^{N}+f_{t}^{N}) +C_{D} (f_{d}^{N}+f_{s}^{N}+f_{b}^{N}) \right ]^2.
\label{sig2}
\end{equation}

\begin{figure}[th]
\begin{center}
\vspace{0.2cm}
\centerline{\epsfxsize=6cm\epsffile{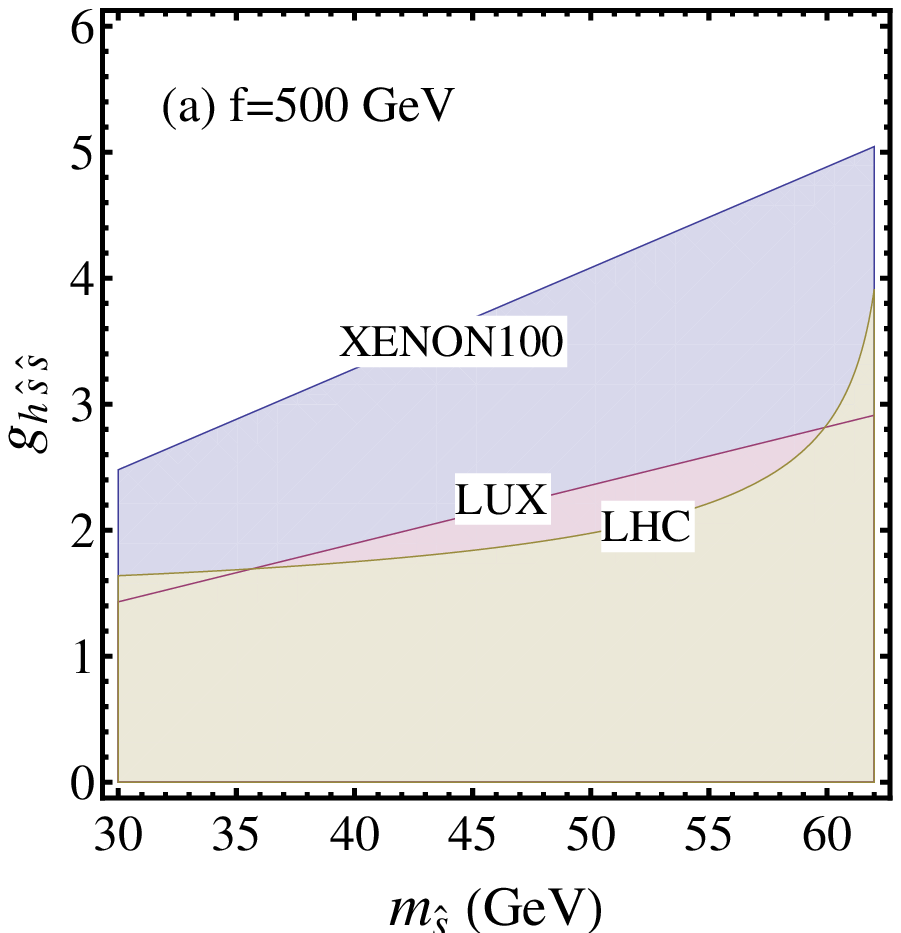}
\hspace{0.5cm}\epsfxsize=6cm\epsffile{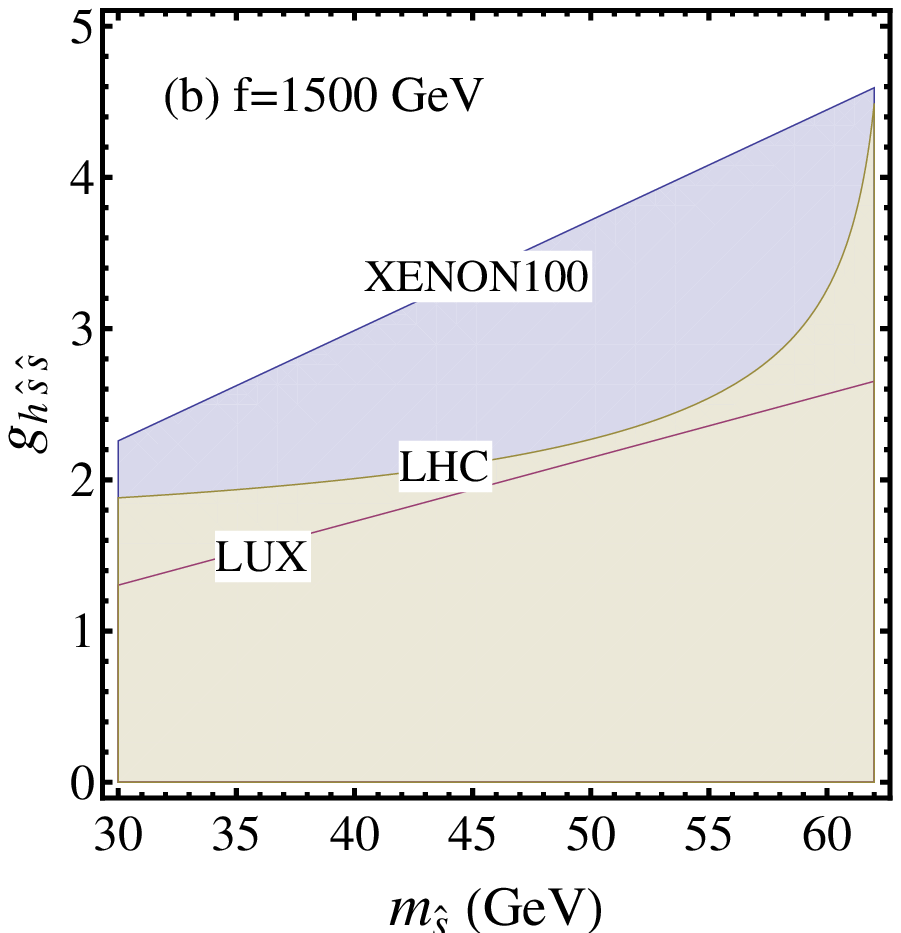}}
\caption{The allowed parameter space in $g_{h\hat{S}\hat{S}}-m_{\hat{S}}$ plane
for $M=150$ GeV and two typical values of $f$:  (a) $f=500$ GeV, and (b) $f=1500$ GeV.
The upper straight line comes from the $90\%$ C.L. upper limits
of XENON100 \cite{xenon}, the lower straight line comes from the $90\%$ C.L. upper
limits of LUX \cite{lux}. The curve come from the LHC Higgs data at 2$\sigma$ level. }
\end{center}
\end{figure}
\begin{figure}[t]
\begin{center}
\vspace{0.2cm}
\centerline{\epsfxsize=6cm\epsffile{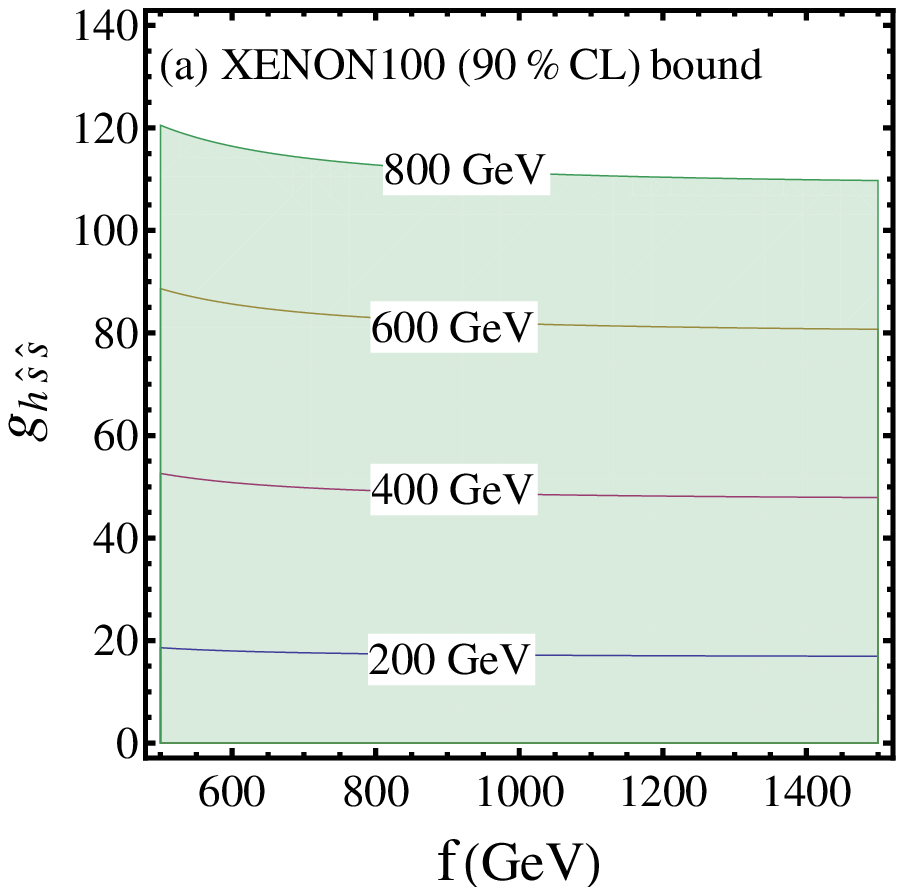}
\hspace{0.5cm}\epsfxsize=6cm\epsffile{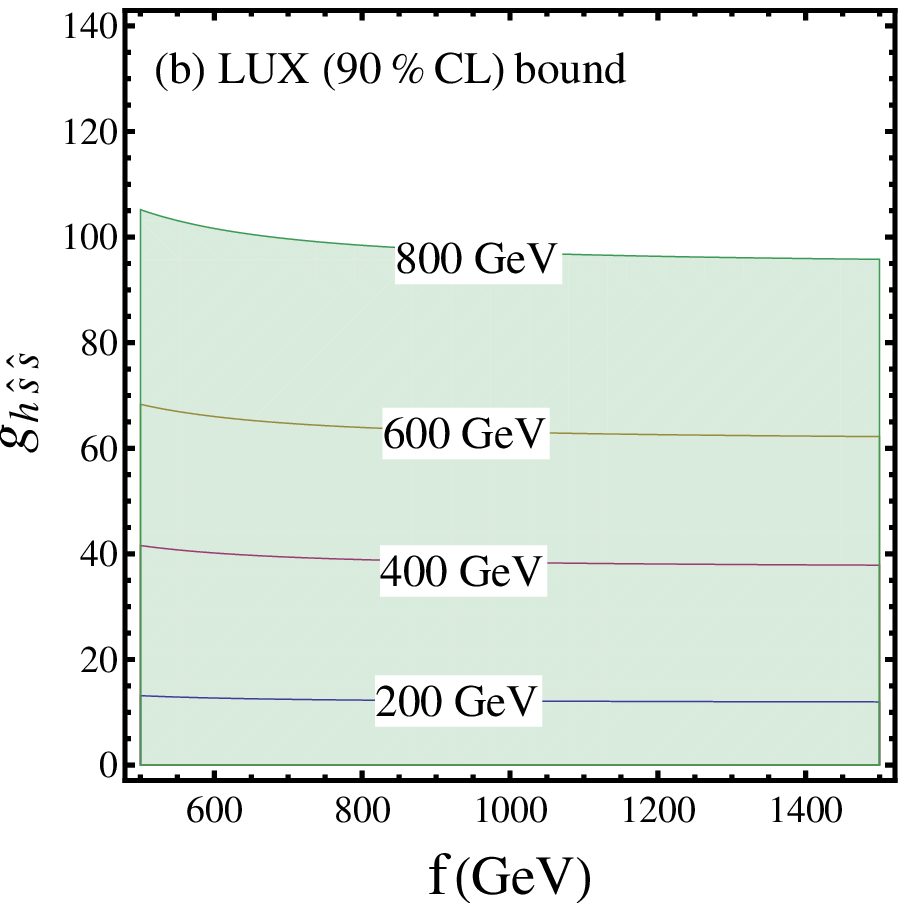}}
\caption{The allowed parameter space in $g_{h\hat{S}\hat{S}}-f$ plane for $M=150$ GeV and
four typical values of DM mass, $m_{\hat{S}}=(200, 400, 600, 800)$ GeV (from bottom to top)
under the  $90\%$ C.L. upper limits of (a) XENON100, and (b) LUX. }
\end{center}
\end{figure}

In Figs.~5, we show the allowed region on DM mass and $g_{h\hat{S}\hat{S}}$ coupling plane which are
consistent with the LHC limit, XENON100 and LUX bounds for $M=150$ GeV and two typical values
of $f=500$ GeV and $f=1500$ GeV.
It is obvious from Fig.~5 that the coupling parameter of $g_{h\hat{S}\hat{S}}$ is insensitive to
the variation of the scale parameter $f$.
In the region of $30$ GeV $\leq m_{\hat{S}}\leq 62.5$ GeV, the value of $g_{h\hat{S}\hat{S}}$ must be
smaller than 5.
On the other hand, we can see that for a large value of scale parameter $f$, the strongest
constraint comes from the LUX upper limit on the DM-nucleon scattering cross section.

In Fig.~6, we display the $g_{h\hat{S}\hat{S}}$ ranges allowed by the upper limits of XENON100 and
LUX on the spin-independent cross section. We can see that the coupling strength of $h\hat{S}\hat{S}$
is sensitive to the DM mass but insensitive to $f$. The results of direct DM searches with LUX
can give strict constraints on the viable parameter space, especially on the DM mass and the
coupling strength between the SM-like Higgs boson and DM pair.
For $m_{\hat{S}}=400$ GeV, $g_{h\hat{S}\hat{S}}$ should be smaller than about 40 according
to the LUX $90\%$ CL data.

In this paper we do not consider the possible relic density bound on the couplings parameter $g_{h\hat{S}\hat{S}}$ in
the LRTH model because it has been systematically studied in Ref.~\cite{sf}, where the authors
identified regions of parameters space that satisfies the correct relic density  requirement.
Specifically, they considered the possible mass splitting between the dark matter $\hat{S}$ and other
particles $\hat{A}$ and $\hat{h}_{1}^{\pm}$, and showed  that the right relic density is rather
sensitive to the mass splitting parameters  $\delta_{1}=m_{\hat{h}_1}-m_{\hat{S}}$ and $\delta_{2}=m_{\hat{A}}-m_{\hat{S}}$.
From the Figs.(3-9) in Ref.~\cite{sf}, we get to know that the combined WMAP and Planck data ($\Omega h^{2}=0.1126\pm0.0036$)
cannot put strong limit on the parameter $ g_{h\hat{S}\hat{S}}$ because the possible limit itself is
sensitive to the variation of new parameters $\delta_{1}$ and $\delta_{2}$.
For more details about this point one can see Ref.~\cite{sf}.

\section{Conclusions}

The LRTH model can provide a scalar boson $\hat{S}$ as a natural candidate for the WIMP
DM. In this paper, we have examined the status of a light dark matter (which open new
decay channels of the SM-like Higgs boson) under current experimental constraints
including the latest LHC Higgs data, the XENON100 and latest LUX limit on the dark
matter scattering off the nucleon.
We also consider the constraints of the DM experimental results on the
parameter space of the coupling strength $g_{h\hat{S}\hat{S}}$ for both low and
high DM mass regions. From the numerical results we obtain the following observations:
\begin{enumerate}
\item
The current ATLAS (CMS) measurements of $R_{\gamma\gamma}$
can exclude the invisible Higgs branching ratio ${\rm Br}_{\rm inv}$ larger than  $34\%$ ($48\%$) at $2\sigma$ level.

\item
The Global fits to the latest LHC and Tevatron Higgs data provide a stronger constraint on the invisible
decay rate: ${\rm Br}_{\rm inv}< 20\%$ ($30\%$) at $2\sigma$ ($3\sigma$) level,
which could be tested at the LHC experiments.

\item
For the spin-independent scattering cross section off the nucleon, the recent XENON100 data can exclude
the invisible decay rate above $50\%$, while the latest LUX result can exclude an invisible
decay branching ratio larger than $25\%$.

\item
The results of direct DM searches with LUX can give strong constraint on the viable  parameter
space of $\{ g_{h\hat{S}\hat{S}},  m_{\hat{S}}\}$ in this LRTH model.
\end{enumerate}

\begin{appendix}

\section{The couplings of $h\hat{h}_{2}\hat{h}_{2}$ and $h\hat{h}^{+}_{1}\hat{h}^{-}_{1}$ arising from the CW potential}\label{sec:couplings}
From \cite{suweb}, the couplings of  $h\hat{h}_{2}\hat{h}_{2}$ and $h\hat{h}^{+}_{1}\hat{h}^{-}_{1}$ arising from the one-loop CW potential can be written as:
\beq
h\hat{h}_{2}\hat{h}_{2}&:& -\frac{3vg_{2}^{2}}{128\pi^{2}}\{ (6+2A1+2A2)g_{2}^{2}+\frac{(2g_{1}^{2}+g_{2}^{2})}{(g_{1}^{2}+g_{2}^{2})^{2}}[(2g_{1}^{2}g_{2}^{2}+g_{2}^{4})A3\\
& &+g_{1}^{4}+6g_{1}^{2}g_{2}^{2}+3g_{2}^{4}+(2g_{1}^{4}+2g_{1}^{2}g_{2}^{2}+g_{2}^{4})A4]\},\\
h\hat{h}^{+}_{1}\hat{h}^{-}_{1}&:& -\frac{3vg_{2}^{4}}{128\pi^{2}}\{6+2A1+2A2+\frac{(3+A3)g_{2}^{4}+6g_{1}^{2}g_{2}^{2}+(5+2A4)g_{1}^{4}}{(g_{1}^{2}+g_{2}^{2})^{2}}\},
\eeq
with
\beq
A1&=&\frac{3}{2}+\ln[\frac{v^{2}g_{2}^{2}}{4\Lambda^{2}}],\quad  \hspace{1.8cm}A2=\frac{3}{2}+\ln[\frac{(f^{2}+\hat{f}^{2})g_{2}^{2}}{4\Lambda^{2}}],\\
A3&=&\frac{3}{2}+\ln[\frac{g_{2}^{2}(2g_{1}^{2}+g_{2}^{2})v^{2}}{4(g_{1}^{2}+g_{2}^{2})\Lambda^{2}}],\quad A4=\frac{3}{2}+\ln[\frac{(g_{1}^{2}+g_{2}^{2})(f^{2}+\hat{f}^{2})}{2\Lambda^{2}}],
\eeq
where $\Lambda=4\pi f$, the gauge couplings $g_{1}$ and $g_{2}$ are related to $e$ and Weinberg angle $\theta_{W}$ and can be written as
\beq
g_{1}=\frac{e}{\sqrt{\cos2\theta_{W}}}, \quad g_{2}=\frac{e}{\sin \theta_{W}}.
\eeq
The values of $f$ and $\hat{f}$ are interconnected once we set $v=246$ GeV.

\section{The Higgs decays $h\to  gg, \gamma\gamma$ }\label{sec:higgs}

In the LRTH model, the leading-order decay widths of $h\to  gg, \gamma\gamma$ are given by
\beq
 \Gamma(h\to gg)&=&\frac{\sqrt{2}G_{F}\alpha_{s}^{2}m_{h}^{3}}{32\pi^{3}}
 \Bigl |-\frac{1}{2}F_{1/2}(\tau_{t})y_{t}y_{G_{F}}-
 \frac{1}{2}F_{1/2}(\tau_{T})y_{T} \Bigr |^{2},\\
 \Gamma(h\to  \gamma\gamma)&=&\frac{\sqrt{2}G_{F}\alpha_{e}^{2}m_{h}^{3}}{256\pi^{3}}
 \Bigl |\frac{4}{3}F_{1/2}(\tau_{t})y_{t}y_{G_{F}}+\frac{4}{3}F_{1/2}(\tau_{T})y_{T}
 +F_{1}(\tau_{W})y_{W}\non
 & & F_{1}(\tau_{W_{H}})y_{W_{H}}+y_{\hat{h}_{1}^{+}\hat{h}_{1}^{-}h}F_{0}(\tau_{\hat{h}_{1}^{\pm}}) \Bigr |^{2},
\eeq
with
\beq
 F_{1}&=&2+3\tau+3\tau(2-\tau)f(\tau),\quad
F_{1/2}=-2\tau[1+(1-\tau)f(\tau)],\quad  F_{0}=\tau[1-\tau f(\tau)], \non
f(\tau)&=&\left [\sin^{-1}(1/\sqrt{\tau}) \right ]^{2},~~~~~~~~~~\quad
g(\tau)= \sqrt{\tau-1}\sin^{-1}(1/\sqrt{\tau}),
\eeq
for $\tau_{i}=4m_{i}^{2}/m_h^2 \geq  1$. The relevant couplings $ y_{i}$ can be written as
\beq
y_{t}= S_{L}S_{R},\quad y_{T}= \frac{m_{t}}{m_{T}}C_{L}C_{R},\quad y_{W_{H}}=-\frac{m^{2}_{W}}{m^{2}_{W_{H}}}, \quad y_{\hat{h}_{1}^{+}\hat{h}_{1}^{-}h}=-\frac{v}{2m_{\hat{h}_{1}}^{2}}g_{\hat{h}_{1}^{+}\hat{h}_{1}^{-}h},
\eeq
where the mixing angles $S_{L,R}$ and $C_{L,R}$ are of the form
\beq
S_{L}&=& \frac{1}{\sqrt{2}}\sqrt{1-(y^{2}f^{2}\cos2x+M^{2})/N_{t}},
\quad C_{L}=\sqrt{1-S_L^2}, \label{eq:sl1}\\
S_{R}&=& \frac{1}{\sqrt{2}}\sqrt{1-(y^{2}f^{2}\cos2x-M^{2})/N_{t}},
\quad C_{R}=\sqrt{1-S_{R}^2},\label{eq:sr1}
\eeq
with
\beq
N_{t}=\sqrt{(M^{2}+y^{2}f^{2})^{2}-y^{4}f^{4}\sin^{2}2x},\label{eq:nt1}
\eeq
with $x=v/(\sqrt{2}f)$. The parameter $M$ is essential to the mixing between the
SM-like top quark and its partner $T$. Here we have neglected the contributions
from $\phi^\pm$ for the $h\to  \gamma\gamma$ decay, this is because their contributions
are even much smaller than that from the $T$-quark \cite{liu-1311}.

The masses of the SM-like top quark and heavy $T$-quark are given by \cite{Hock}
\beq
m_{t}^{2}= \frac{1}{2}(M^{2}+y^{2}f^{2}-N_{t}), \quad m_{T}^{2}= \frac{1}{2}(M^{2}+y^{2}f^{2}+N_{t}),
\eeq
with the parameter $N_t$ as defined in Eq.~(\ref{eq:nt1}).

Besides the SM top, $W^\pm$, $T$-quark and $W_H$, the light scalar $\hat{h}_{1}^{\pm}$ can also contribute to the
Higgs diphoton rate, but such contribution is indeed very small in size.
In Table II, we list the relative strength of the possible contributions to the decay width
$\Gamma(h\to  \gamma \gamma)$ coming from different sources.
One can see from the numbers in Table II that, the dominant contribution
do come from the SM top and $W^\pm$, the new physics contributions from $W_H$ and $\hat{h}_{1}$
are even much smaller than that from the $T$-quark, and can be neglected safely.

\begin{table}[thb]
\begin{center}
\caption{ The relative strength of the contributions to the decay
amplitude of $h\to \gamma\gamma$ in the LRTH model from different sources,
assuming  $M=150$ GeV and $f=600$ GeV, $m_{\hat{h}_{1}}=60$ GeV, 100 GeV and 400 GeV, respectively. }
\label{table2} \vspace{0.2cm}
\begin{tabular}{|c|c|c|c|c|c|c|}  \hline $m_{\hat{h}_{1}}$&SM top&$W^\pm$&$T$-quark&$W_{H}$&$\hat{h}_{1}$& Total \\ \hline
60 GeV&-1.76&7.98&0.14&-0.022&0.055&6.39\\
\hline 100 GeV&-1.76&7.98&0.14&-0.008&0.008&6.35\\
\hline 400 GeV&-1.76&7.98&0.14&-0.022&0.0004&6.34\\
\hline  \end{tabular}\end {center} \end{table}

\end{appendix}

\begin{acknowledgments}
Yao-Bei Liu thank Lei Feng for helpful discussion on MicrOmergas. We also thank Shufang Su for helpful discussions.
This work is supported by the National Natural Science
Foundation of China under the Grant No. 11235005, the Joint Funds
of the National Natural Science Foundation of China (U1304112), and by the Project on Graduate Students Education
and Innovation of Jiangsu Province under Grant No. KYZZ-0210.

\end{acknowledgments}


\end{document}